\documentclass[final,5p,times,twocolumn]{elsarticle}

\usepackage{xspace}
\usepackage{xcolor}
\usepackage{comment}
\usepackage{adjustbox}
\usepackage{numprint}
\npthousandsep{,}

\newcommand{\urltt}[1]{\textbf{\texttt{\url{#1}}}}

\newcommand{\ie}{\emph{i.e.,}\xspace}
\newcommand{\eg}{\emph{e.g.,}\xspace}
\newcommand{\etc}{etc.\xspace}
\newcommand{\etal}{\emph{et~al.}\xspace}
\newcommand{\aka}{\emph{a.k.a.}\xspace}

\newcommand{\seesection}[1]{(see~\secref{#1})}
\newcommand{\secref}[1]{Section~\ref{#1}\xspace}

\newcommand{\figref}[1]{Figure~\ref{#1}\xspace}

\newcommand{\tabref}[1]{Table~\ref{#1}\xspace}

\newcommand{\InsightBox}[1]{\noindent\framebox[\columnwidth][l]{\quad\parbox[h]{0.89\columnwidth}{ #1 }}}

\newcommand*\circledInverted[1]{\tikz[baseline=(char.base)]{\node[fill=black,text=white,shape=circle,draw,inner sep=0.8pt] (char) {\small{}#1};}}

\usepackage{enumitem}
\usepackage{fontawesome5}
\usepackage{xurl}
\usepackage{hyperref}
\usepackage{booktabs}
\usepackage{tikz}
\usepackage{multibib}

\newcites{white}{References}
\newcites{grey}{Web references}

\begin{document}
\title{An Empirical Study on Database Usage in Microservices} 


\affiliation[1]{organization={University of Namur},
  country={Belgium}}

\affiliation[2]{organization={REVEAL @ Software Institute --- USI, Lugano}, country={Switzerland}}

\affiliation[3]{organization={University of Zurich},
  country={Switzerland}}

\author[1]{Maxime André\corref{cor1}} 
\ead{maxime.andre@unamur.be}

\author[2]{Marco Raglianti}
\ead{marco.raglianti@usi.ch}

\author[3]{Souhaila Serbout}
\ead{souhaila.serbout@uzh.ch}

\author[1]{Anthony Cleve}
\ead{anthony.cleve@unamur.be}

\author[2]{Michele Lanza}
\ead{michele.lanza@usi.ch}

\cortext[cor1]{Corresponding author}

\begin{abstract}
Microservices architectures are an integral part of modern software development. Their adoption brings significant changes to database management. Instead of relying on a single database, a microservices architecture is typically composed of multiple, smaller, heterogeneous, and distributed DBs. In these data-intensive systems, the variety and combination of database categories and technologies play a crucial role in storing and managing data. While data management in microservices is a major challenge, research literature is scarce.

We present an empirical study on how databases are used in microservices. On the dataset we collected (and released as open data for future research), considering 15 years of microservices, we examine ca. 1,000 GitHub projects that use databases selected among 180 technologies from 14 categories. We perform a comprehensive analysis of current practices, providing researchers and practitioners with empirical evidence to better understand database usage in microservices. We report 18 findings and 9 recommendations. We show that microservices predominantly use Relational, Key-Value, Document, and Search databases. Notably, 52\% of microservices combine multiple database categories. Complexity correlates with database count, with older systems favoring Relational databases and newer ones increasingly adopting Key-Value and Document technologies. {\em Niche} databases (\eg EventStoreDB, PostGIS), while not widespread, are often combined with a {\em mainstream} one.
\end{abstract}


\maketitle

\newpage


\section{Introduction}

Microservices architectures have significantly gained popularity, becoming an integral part of the software development landscape. This architectural style is now widely adopted by large and software-intensive companies like Amazon, Google, and Netflix~\citewhite{richardson2018,newman2021}.

Their adoption brings significant changes to database (DB) management~\citewhite{laigner2021,seattle2022,andre2023,assunccao2023}. According to the literature, the microservices architecture paradigm, which promotes the decomposition of a system into loosely coupled, independent, heterogenous, and manageable services, is expected to naturally extend to DBs~\citewhite{assunccao2023}. Specifically, each microservice is expected to have its own dedicated DB(s) following the \textit{database server per microservice} pattern~\citewhite{laigner2021}, ensuring data autonomy and minimizing dependencies. This is aligned with the \textit{polyglot persistence}~\citegrey{fowler2011,lewis2014}. In these data-intensive systems, the large variety and combination of DB categories and technologies play a crucial role in storing and managing data. Depending on the requirements, the main motivations concern, for instance, the need for independent schema evolution, data caching, data replication, data partitioning, decentralized data management, \etc These mechanisms aim to reduce coupling and ease maintenance and evolution~\citewhite{laigner2021}.

Although a few existing studies recognize that data management in microservices is a major challenge~\citewhite{furda2018,laigner2021}, it received little attention in the research literature. Current works lack concreteness, especially regarding the available datasets and in-depth empirical investigations that highlight the current status. In particular, the variety and numerous combinations of DB categories and the specific technologies and their implementations, across multiple heterogeneous microservices, often require precise justifications to understand the underlying reasons emerging from the community trends. Indeed, despite the growing adoption of microservices architectures, there is still a noticeable gap and a lack of benchmarks in the literature~\citewhite{laigner2021} regarding how microservices practitioners reason and handle data management {\em in vivo}. Existing studies~\citewhite{laigner2021,benats2021,assunccao2023,paiva2025,graetsch2025} confirm a trend in the adoption of multiple DBs in modern software, such as the combination of relational and document DBs, the use of a cache layer, and the exploitation of search-based mechanisms. Some conclude that a poor understanding of data management practices, such as technology combinations, could lead to the introduction of a \textit{technical data debt}~\citewhite{graetsch2025}. To fill this gap, concrete observations on the state of the practice could be useful to practitioners, teachers, and students, helping them to make the right technical choices.

We present an empirical study on how DBs are used in microservices. Considering 15 years of history, from 2010 to 2025, we examine ca. \numprint{1000} open-source projects mined from GitHub that use DBs selected among 180 technologies (\eg PostgreSQL, Redis, MongoDB) from 14 categories (\eg Relational, Key-Value, Document).

\newpage

We perform a comprehensive analysis providing insights into current practices, emphasizing the most prevalent DBs used in microservices. We also investigate the way they are combined in practice, observing recurrent patterns. We support our observations with objective, fine-grained metrics and highlight relationships between characteristics (\eg complexity vs. age).

Our study leads to 18 findings about DB usage in microservices, with a two-fold implication. First, on the industry and open-source side, our work helps practitioners to understand the latest trends and, thanks to the 9 recommendations we derive, to select the most appropriate data storage strategies in their projects. Second, on the research side, it guides researchers in shaping future directions based on empirical evidence and an open-source dataset.

\section{Research Methods} \label{section:methods}

We describe the research methods we employed to conduct our empirical study. We present our research question, the initial list of considered DBs and technologies, and the methodology we followed to collect and analyze the data from GitHub repositories.

\subsection{Research Questions} \label{section:research_questions}

We aim to answer the following research questions (noted \textbf{RQ*}) to understand how DBs are used in microservices:

\begin{itemize}[leftmargin=5mm]

\item \textit{\textbf{RQ1}: What database categories and technologies are used in microservices, and how prevalent are they?} From the open-source microservices collected, we analyze the DB dependencies, establish their distribution, and assess the most popular DB categories (\eg Relational, Document, Key-Value, Column, Graph) and technologies (\eg PostgreSQL, MongoDB, Redis).

\item \textit{\textbf{RQ2}: How are databases combined in microservices, and what are the characteristics of those combinations?} We analyze the DB categories and technologies associations in microservices, their breakdown as stated in dependencies, and determine the most popular combinations, exploring further recurrent patterns and computing relevant metrics.

\item \textit{\textbf{RQ3}: What is the relationship between the complexity of microservices and their data management strategy?} We seek to understand whether the complexity (\eg number of services, size of the project) is correlated with the number of DB technologies, whether the complexity is linked to a certain degree of category associations, or whether some category associations are more suitable for projects of certain complexity.

\item \textit{\textbf{RQ4}: What is the relationship between the age of microservices and their database choices?} We consider the age of microservices and aim to find rationales in their DB choices, to recommend different strategies for older and more recent projects.

\end{itemize}

\subsection{Database Categories and Technologies} \label{section:databases}

\tabref{table:database_categories} lists the DB categories considered in our study. We extracted them from DB-Engines' March 2025 ranking~\citegrey{DBEngines}, considering the top 250 DB management systems (DBMSs). The exhaustive list of DB technologies is available in our replication package (\secref{section:replication_package_dataset}).

\begin{table}[ht]
  \centering
  {\small
  \caption{DBMS categories considered in our study.}
    \label{table:database_categories}
  \begin{tabular}{l|l} \toprule
      \textbf{Category} & \textbf{Example DBMS} \\ \midrule
      Relational      & Oracle, MySQL, MS SQL Server, PostgreSQL\\
      Document        & MongoDB, Couchbase, CouchDB \\
      Key-Value       & Redis, Memcached, etcd \\ 
      Column          & Cassandra, HBase, ClickHouse \\
      Graph           & Neo4j, GraphDB \\
      Time Series     & InfluxDB, kdb+, TimescaleDB \\
      Vector          & Pinecone, Milvus, Qdrant, Chroma, Weaviate \\
      Spatial         & PostGIS \\
      Hierarchical    & IBM IMS \\
      Network         & IDMS \\
      Object          & Actian, db4o, ObjectDB \\
      Event           & EventStoreDB \\
      Search          & Elasticsearch, Splunk, Solr \\
      Others          & Amazon DynamoDB, Aerospike \\ \bottomrule
  \end{tabular}}

\end{table}

Since the presence as a Docker container is a distinction criterion \seesection{section:github}, we excluded DB technologies unavailable as Docker images (\eg Microsoft Access, FileMaker). Following the methodology of Paiva \etal, we also excluded warehouses (\eg Snowflake, Databricks, Apache Hive, Google BigQuery), frameworks (\eg Apache Flink), services (\eg Amazon Aurora, Prometheus), and platforms (\eg Google Firebase, Google Firestore, Microsoft Azure Table Store)~\citewhite{paiva2025}. In the end, we considered a total of 180 technologies.

DBs are categorized by type (Relational, Document, Key-Value, Column, Graph, Time Series, Vector, Spatial, Hierarchical, Network, Object, Event, Search Engine) and ranked by popularity according to DB-Engines. Information and classifications were cross-verified through the literature~\citewhite{viennot2015,gessert2017,davoudian2018,gan2018,gan2019,laigner2021,benats2021,paiva2025} and online sources such as Database of Databases~\citegrey{dbdb} and Wikipedia~\citegrey{wikiRelationalDatabases,wikiObjectDatabases,wikiDocumentDatabases,wikiColumnDatabases,wikiMemoryDatabases,wikiNoSQL}.

For each DB, the first category is considered to be the main one. Less common, ambiguous, or multi-type DBs (\eg Native XML, RDF), are grouped under the ``Others'' category. Each DB is associated with a regular expression (RE) to identify its corresponding Docker image, if available. REs, verified on Docker Hub~\citegrey{dockerhub}, are included in our replication package.

\begin{figure*}[ht]
  \centering
    \includegraphics[width=0.6\linewidth]{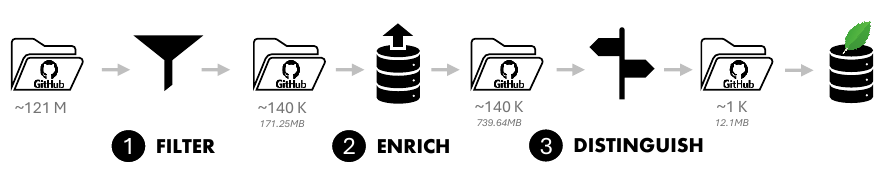}
    \caption{\label{figure:mining}The mining process for extracting microservices with DBs from GitHub.}
\end{figure*}

\subsection{GitHub Repositories and Microservices} \label{section:github}

We mined GitHub using its REST (REpresentational State Transfer) API (Application Programming Interface)~\citegrey{GitHubRESTAPI} to collect microservices project repositories that use DBs. Mining repositories is valuable for researchers seeking up-to-date real-world systems to evaluate their approaches and tools~\citewhite{kalliamvakou2014}. Benchmarks and datasets specifically studying microservices and DBs are limited in the literature~\citewhite{rahman2019,dAragona2024}. There are several challenges to be addressed to obtain a good benchmark.

First, it is not straightforward to identify a GitHub repository belonging to a system that adheres to a microservices architecture, as they can follow various organizational structures, such as \textit{mono-repositories} (\aka \textit{mono-repo})~\citewhite{assunccao2023} or \textit{multi-repositories} (\aka \textit{multi-repo}). Additionally, documentation that lists and describes all the microservices within a given architecture is often unavailable, further complicating the mining process. When examining a specific repository, there is rarely a clear and explicit indication that it belongs to a microservices architecture. Sometimes, terms like ``microservice'' might help, but other associated terms like ``REST API'' can also be observed in titles, tags, descriptions, or README files, either at the top level of the architecture or at sub-levels representing parts of the architecture. Since a microservice is modular and distributed, it can be difficult to scope. Some components can be spread, isolated in other locations, without any clear links. The heterogeneity of implementations affects the automation capabilities of mining such repositories. These challenges often lead to noisy or incomplete results. Current benchmarks commonly require manual annotation, which slows down the process and limits the number of results included and analyzed. To address these challenges, we propose a mining process that combines several fine-grained filters and heuristics to reduce the difficulty of characterizing microservices repositories and provide a large benchmark dataset of microservices architectures. 

The mining process is depicted in \figref{figure:mining}. To ensure the reproducibility of our research, the source code and the complete benchmark dataset are available in our replication package.

\noindent\textbf{\circledInverted{1} Filter}. Since GitHub may host private and inactive projects~\citewhite{kalliamvakou2014,benats2021,paiva2025}, we applied six filtering criteria to efficiently narrow down our search from 121 million repositories. The aim was to identify the relevant ones that are active real-world systems. After filtering we retained \numprint{140000} repositories based on:

\begin{enumerate}[leftmargin=5mm]
  
\item \textbf{Disk Size:} To eliminate outliers and retain repositories with meaningful content, we filtered them based on disk size~\citewhite{benats2021}. We selected those with a size between 500 KB and 1 GB.

\item \textbf{Stars Count:} To retain relevant repositories by popularity, we applied a filtering criterion based on the number of stars~\citewhite{borges2018}. We set the threshold to at least 100 stars.

\item \textbf{Commit History:} To target real and actively maintained systems, we retained only repositories with at least 100 commits, inspired by the work of d'Aragona \etal~\citewhite{dAragona2024}.

\item \textbf{Structural Completeness:} To remove placeholders and projects without the basic level of documentation encouraged on GitHub, we included only repositories with at least one \textit{README} file and at least two directories~\citewhite{kalliamvakou2014}.

\item \textbf{Recent Updates:} We focused on repositories that are likely to follow modern microservices practices by selecting only those updated after January 1, 2015. This date is often considered to mark the widespread adoption of microservices architectures~\citegrey{lewis2014}.

\item \textbf{Programming Languages:} We targeted repositories written in popular programming languages, particularly those frequently used in microservices development~\citewhite{gan2018,gan2019,rahman2019,dAragona2024}, \citegrey{octoverse2024}. The selected languages include C, C++, C\#, Go, Java, JavaScript, PHP, Python, Ruby, Scala, and TypeScript. 

\end{enumerate}

\noindent\textbf{\circledInverted{2} Enrich}. To better identify microservices repositories, we enrich the dataset with additional information. Using the GitHub API, we retrieve the content of README files, which often contain keywords and valuable project details, as well as \textit{Docker Compose} files,\footnote{\urltt{https://docs.docker.com/compose/}} commonly used to define multi-container environments typical of microservices architectures~\citewhite{assunccao2023,dAragona2024}.

\noindent\textbf{\circledInverted{3} Distinguish}. Based on the enriched dataset, we deepen the filtering process targeting repositories likely to be microservices. We define a number of heuristics to compute a score for each repository. The higher the score, the more likely the repository is a microservice. The heuristics are based on the presence of keywords in the repository, the number of specific files and directories, the presence and content of README and Docker Compose files, and the number of services and DBs declared in the Docker Compose files. 

The keywords are the following variants: \textit{microservice}, \textit{micro-service}, \textit{micro service}, \textit{microservices}, \textit{micro-services}, \textit{micro services}. In addition, we add keyword variants about the different organizational structures: \textit{monorepo}, \textit{mono-repo}, \textit{multirepo}, \textit{multi-repo}. Finally, the \textit{rest api} keyword aims to include the parts of microservices architectures that are outside the scope of a single repository and are intended to be served as external APIs for other services.

\begin{figure*}[ht]
  \centering
  \includegraphics[width=\linewidth]{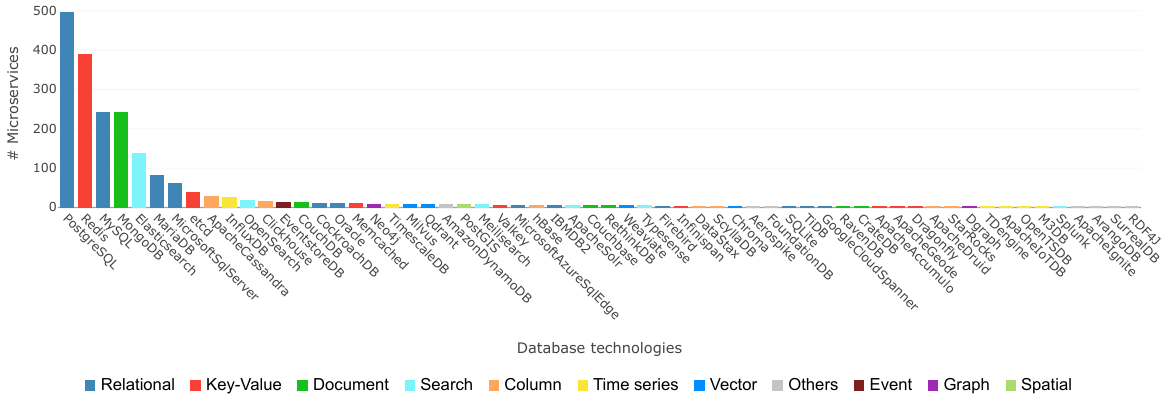}
  \caption{\label{figure:microservices_databases}DB technologies utilized in microservices, colored by category.}
\end{figure*}

\begin{figure*}[ht]
  \centering
  \includegraphics[width=\linewidth]{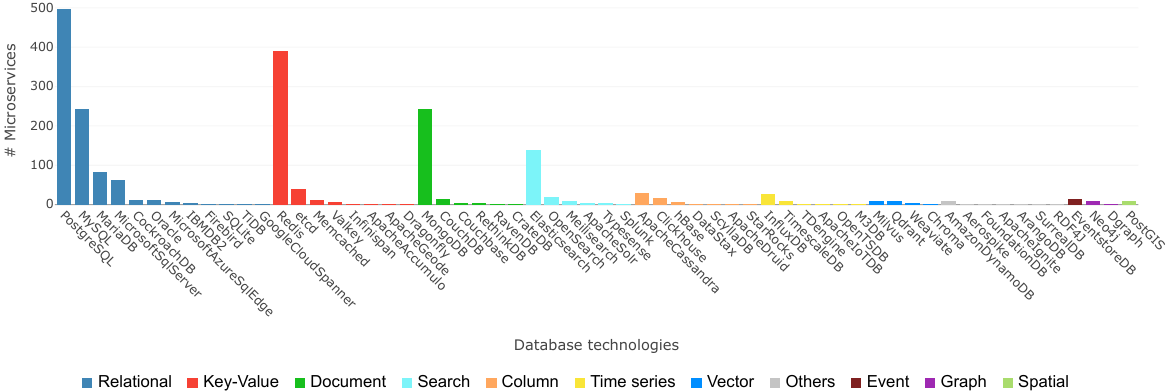}
  \caption{\label{figure:microservices_databases_group_by_categories}DB technologies utilized in microservices, colored and grouped by category.}
\end{figure*}

We compute the score based on the following heuristics:

\begin{itemize}[leftmargin=5mm]
\item A keyword is featured in the title, the description, the repository topics, the contents of the README files.
\item A keyword is featured in at least a directory, a file, a Docker Compose file.
\item The repository Docker Compose files declare at least one service or one DB.
\item The repository counts more services than DBs.
\end{itemize}

These heuristics aim to compute a likelihood score for a repository to have a microservices architecture. All heuristics are assessed independently.

Once the score is computed, in order to flag the repository as a microservices architecture, it must satisfy all of the following conditions:

\begin{itemize}[leftmargin=5mm]
  \item The score is (strictly) greater than 0.
  \item The Docker Compose files declare at least one service.
  \item The repository contains at least one keyword in the title, description, topics, or README files.
  \item The Docker Compose files declare at least one DB.
  \item The repository counts more services than DBs.
\end{itemize}

We found \numprint{1005} repositories that are likely to be microservices. We stored them in a DB containing, for each repository: GitHub ID, GitHub URL, git branch, owner username, repository title, repository description, GitHub associated topics, creation date, last updated date, disk size, star count, commit count, contributor count, directory count, service directories list, service files list, README files and their content, Docker Compose files and their content, service count based on the Docker Compose files, DB list based on the ones declared in the Docker Compose files, and programming languages list. The DB dump is available in our replication package.

\section{Results} \label{section:results}

\subsection{RQ1: Database Usage in Microservices} \label{section:result_rq1}

We compute the distribution of DB technologies and categories across all repositories. For each technology, we count the number of repositories declaring it in their Docker Compose files. In \figref{figure:microservices_databases}, we present the DB technologies (x-axis) we found in our dataset, sorted by popularity (\ie number of repositories that use them, y-axis). In \figref{figure:microservices_databases_group_by_categories}, we grouped them by DB category.

Among the \numprint{1005} repositories in our dataset, a total of 60 distinct DB technologies are identified out of the 180 considered, highlighting significant heterogeneity with 11 different DB categories.

The most popular DB categories are Relational, Key-Value, Document, and Search. As reported in \tabref{table:database_categories_distribution}, Relational DBs appear in 71.64\% of repositories, followed by Key-Value DBs in 42.09\%, Document DBs in 25.77\%, and Search DBs in 16.32\%.

\begin{table}[ht]
  \centering
  {\small
  \caption{Distribution of DB categories in microservices.\label{table:database_categories_distribution}}
  \begin{tabular}{lrr|lrr}
    \toprule
    Category & Count & \% & Category & Count & \%\\
    \midrule
    Relational & 720 & 71.64\% & Time Series & 38 & 3.78\%\\
    Key-Value & 423 & 42.09\% & Vector & 19 & 1.89\%\\
    Document & 259 & 25.77\% & Event & 15 & 1.49\%\\
    Search & 164 & 16.32\% & Graph & 10 & 1.00\%\\
    Column & 50 & 4.98\% & Spatial & 8 & 0.80\%\\
    &&&{\em Others} & 14 & 1.39\%\\
    \bottomrule
\end{tabular}}
\end{table}

A repository is counted in a category if it has at least one technology from it. Relational DBs also dominate in terms of the number of distinct technologies (12), followed by Key-Value DBs (8), Document DBs (6), and Search DBs (6). At the other side of the spectrum we find the Spatial DBs category, with a single technology (PostGIS).

Among our collected data, some categories and technologies cannot be found. Notably, no Hierarchical, Network, or Object DBs are identified in microservices. \tabref{table:database_technologies_distribution} shows the distribution of the top 10 most popular DB technologies.

\begin{table}[ht]
  \centering
  {\small
  \caption{Top 10 DB technologies used in microservices.\label{table:database_technologies_distribution}}
  \begin{tabular}{llrr}
    \toprule
    Technology & Category & Count & Percentage\\
    \midrule
    PostgreSQL & Relational & 498 & 49.55\%\\
    Redis & Key-Value & 390 & 38.81\%\\
    MySQL & Relational & 242 & 24.08\%\\
    MongoDB & Document & 242 & 24.08\%\\
    Elasticsearch & Search & 139 & 13.83\%\\
    MariaDB & Relational & 82 & 8.16\%\\
    Microsoft Sql Server & Relational & 63 & 6.27\%\\
    etcd & Key-Value & 39 & 3.88\%\\
    Apache Cassandra & Column & 30 & 2.99\%\\
    InfluxDB & Time Series & 27 & 2.69\%\\
    \bottomrule
\end{tabular}}
\end{table}

For the Relational category, PostgreSQL and MySQL dominate, with 49.55\% and 24.08\% of repositories, respectively. Redis, for Key-Value DBs, is present in 38.81\% of repositories. MongoDB brings the top Document DB technology on par with the second most popular in the Relational category, with 24.08\%.

\tabref{table:database_category_top1} summarizes the most popular DB technology of each category and its in-category percentage (\ie popularity within the category).

\begin{table}[ht]
  \centering
  {\small
  \caption{Most popular DB technology for each category.\label{table:database_category_top1}}
  \begin{tabular}{llrr}
    \toprule
    Category & Technology & Count & Percentage\\
    \midrule
    Relational & PostgreSQL & 498 & 69.17\%\\ 
    Key-Value & Redis & 390 & 92.20\%\\ 
    Document & MongoDB & 242 & 93.44\%\\ 
    Search & Elasticsearch & 139 & 84.76\%\\ 
    Column & Apache Cassandra & 30 & 60.00\%\\ 
    Time Series & InfluxDB & 27 & 71.05\%\\ 
    Event & EventStoreDB & 15 & 100.00\%\\ 
    Graph & Neo4j & 10 & 100.00\%\\ 
    Vector & Milvus & 9 & 47.37\%\\ 
    Spatial & PostGIS & 8 & 100.00\%\\ 
    \bottomrule
\end{tabular}}
\end{table}

PostgreSQL is chosen 69.17\% of the times when Relational DBs are needed. With MySQL, they almost have a monopoly on Relational DBs. In Key-Value and Document DBs, Redis and MongoDB cover 92.20\% and 93.44\% respectively, making them almost a {\em de facto} standard, save for a few exceptions. Elasticsearch occupies 84.76\% of the Search landscape. Apache Cassandra concerns 60.00\% of Column DBs and, for Time Series DBs, InfluxDB is present in 71.05\% of cases. They are popular alternatives but not the only option in their respective categories. On the contrary, while EventStoreDB, Neo4j, and PostGIS are not widespread, they have almost no alternatives in their respective categories. Finally, Milvus and Qdrant are similarly popular alternatives for Vector DBs.

\medskip
\InsightBox{
  \smallskip
	\textbf{\emph{RQ1: Findings}}
	\smallskip
  \begin{itemize}
    \item[F1.1] \emph{Four different categories of technologies are widespread in microservices architectures (Relational, Key-Value, Document, and Search).}
    \item[F1.2] \emph{Relational DBs are still prevalent, by a large margin, with multiple popular alternative technologies.}
    \item[F1.3] \emph{Specific DB categories, despite their niche applications, should not be overlooked. For example, Graph and Spatial DBs, while addressing specialized needs, are utilized in 10 and 8 repositories respectively, indicating their relevance.}
    \item[F1.4] \emph{For most categories, there is a specific technology which is the most widespread (\eg Redis for Key-Value, MongoDB for Document). Relational DBs have a more varied ``footprint'' of popular technologies.}
    \item[F1.5] \emph{Sixty unique DB technologies show that diversity is preferred to one-size-fits-all DBs.}
  \end{itemize}
}

\subsection{RQ2: Database Associations in Microservices} \label{section:result_rq2}

\begin{figure*}[!ht]
  \centering
  \includegraphics[width=1\linewidth]{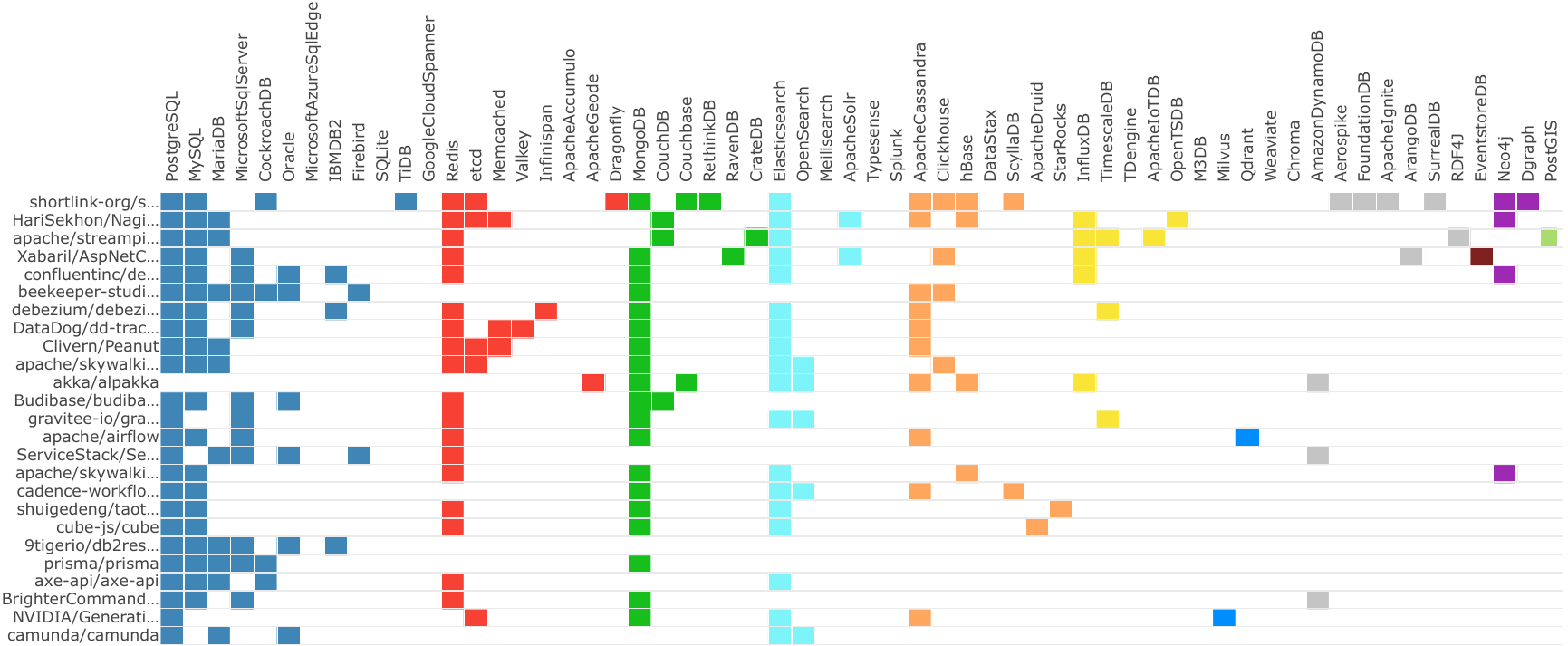}
  \caption{\label{figure:databases_technologies_microservices_subset}Top 25 repositories sorted by number of DB technologies, colored by categories.}
\end{figure*}

For each repository and each Docker Compose file, we inspect the declared DB technologies. \figref{figure:databases_technologies_microservices_subset} presents the repositories (rows) and the DB technologies used (columns). The repositories are sorted by number of DB technologies and the DBs by popularity (see RQ1). The cells are filled if the repository declares the DB technology in one of its Docker Compose files. Only the top 25 repositories are shown. A full interactive version, allowing access to the DB declaration in GitHub, is provided in the replication package.

\begin{figure*}[!ht]
  \centering
  \includegraphics[width=1\linewidth]{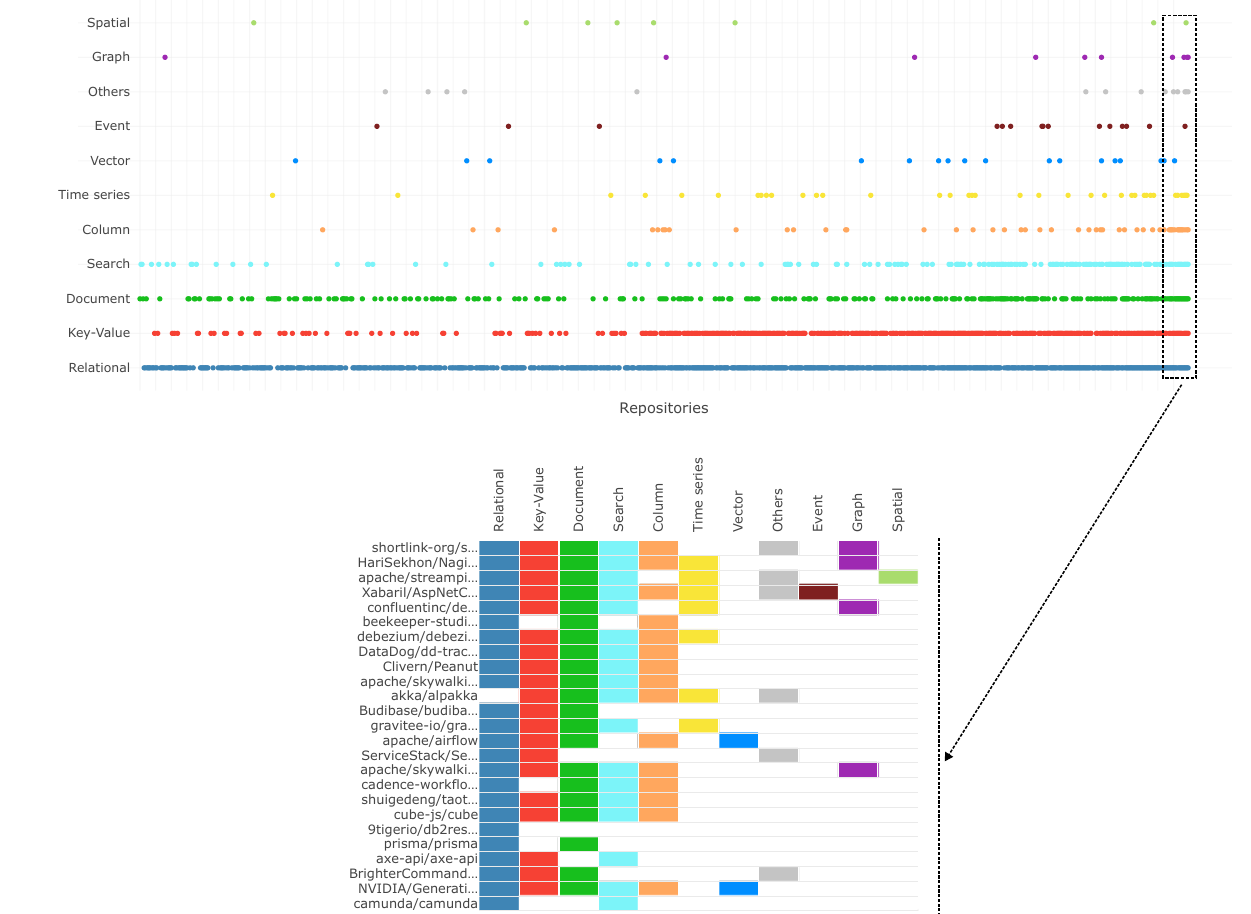}
  \caption{\label{figure:databases_technologies_microservices_group_by_categories_overview}Repositories and their DB categories overview (top). Zoom-in on the top 25 repositories sorted by number of DB technologies (bottom).}
\end{figure*}

\figref{figure:databases_technologies_microservices_group_by_categories_overview} (top) proposes an overview of the \numprint{1005} repositories (columns) and their DB categories (rows).

The sorting of repositories and DBs remain the same. The colored dots show the presence of the DB category in the repository. A zoom-in is done on the top 25 repositories in \figref{figure:databases_technologies_microservices_group_by_categories_overview}, bottom. A full version is available in the replication package. 

The aim of these figures is to demonstrate the heterogeneity of DBs in microservices and to highlight the most popular ones. On the \numprint{1005} repositories collected, we compute what we call the \textit{database heterogeneity rate} (DHR), which is the ratio between the number of microservices combining at least two DB technologies (or categories) and the total number of microservices. The DHR based on technologies is 0.52: Half of the repositories mix two different technologies. The DHR based on categories is 0.47, which highlights that some microservices also use different technologies within the same category.

\begin{figure}[!ht]
  \centering
  \includegraphics[width=\linewidth]{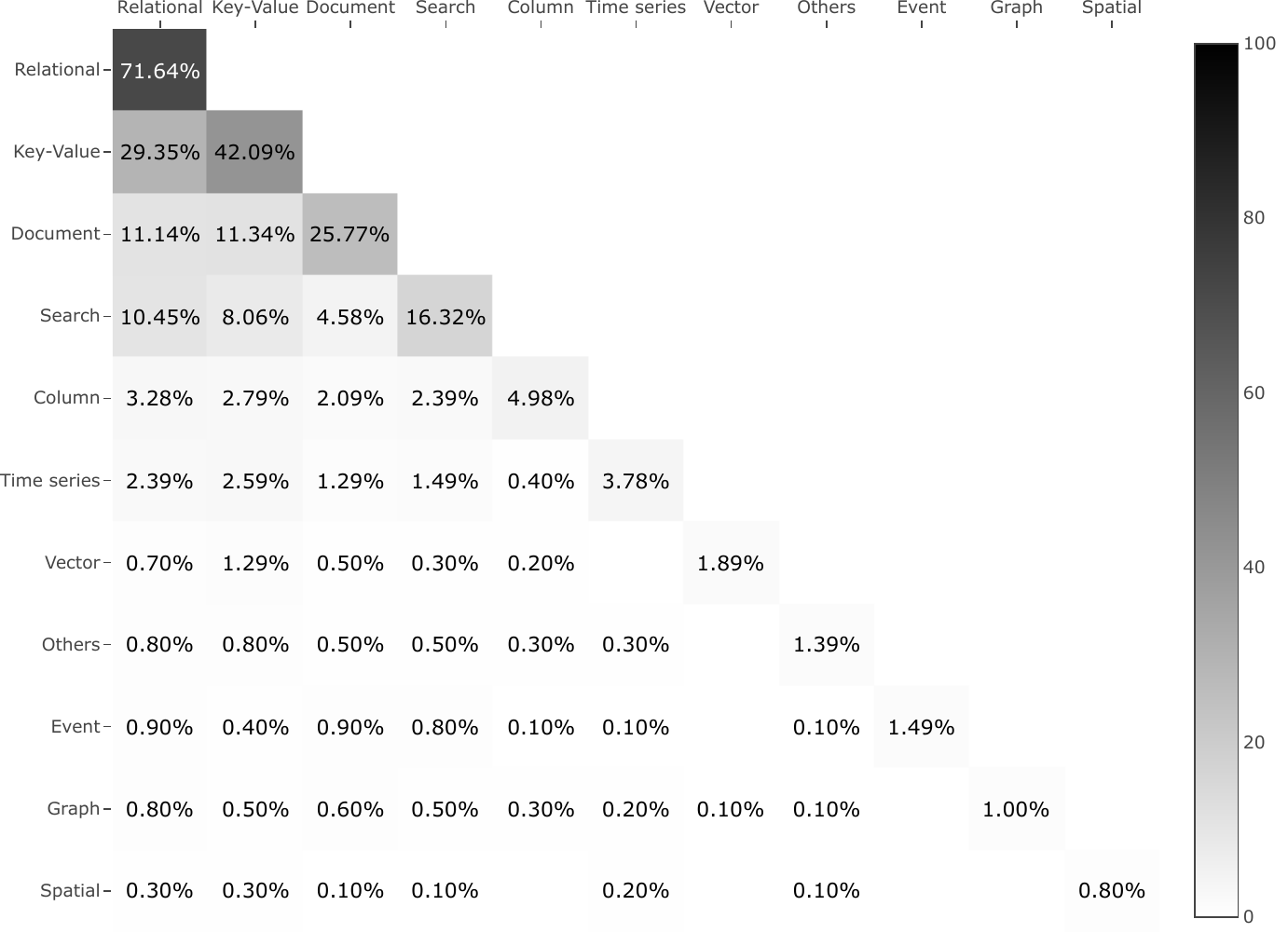}
  \caption{\label{figure:databases_categories_dual_associations}Pairwise combinations of DB categories.}
\end{figure}

In \figref{figure:databases_categories_dual_associations}, we present a cross-matrix showing, for each pair of possible DB categories, the percentage of combinations. The y-axis and x-axis represent DB categories. They are sorted by popularity according to the results obtained previously. Cells are filled with a color gradient from white to black depending on the percentage. Cells account also for the combinations of more than just the exclusive pair, as long as the two categories are present in the tuple. The diagonal indicates the percentage of repositories having the category.

$\langle$Relational, Key-Value$\rangle$ is the pair of DB categories most frequently combined, in 29.35\% of cases, followed by $\langle$Key-Value, Document$\rangle$ in 11.34\% of cases, and closely by $\langle$Relational, Docu\-ment$\rangle$ in 11.14\% of cases. Combinations paired with Search are quite popular: 10.45\% for $\langle$Relational, Search$\rangle$, 8.06\% for $\langle$Key-Value, Search$\rangle$, and 4.58\% for $\langle$Document, Search$\rangle$. The remaining associations appear in less than 3\% of cases. Some are non-existent: No combinations appear for $\langle$Column, Spatial$\rangle$, $\langle$Time Series, Vector$\rangle$, $\langle$Vector, Event$\rangle$, $\langle$Vector, Spatial$\rangle$, $\langle$Event, Graph$\rangle$, $\langle$Event, Spatial$\rangle$, or $\langle$Graph, Spatial$\rangle$, suggesting that niche categories are generally not associated with each other. To overcome the limited overview provided by pairwise associations, we report all combination patterns observed for the top 5 DB categories with all the subsets in their power set (and exclusive use).

\begin{figure}[!ht]
  \centering
  \includegraphics[width=0.85\linewidth]{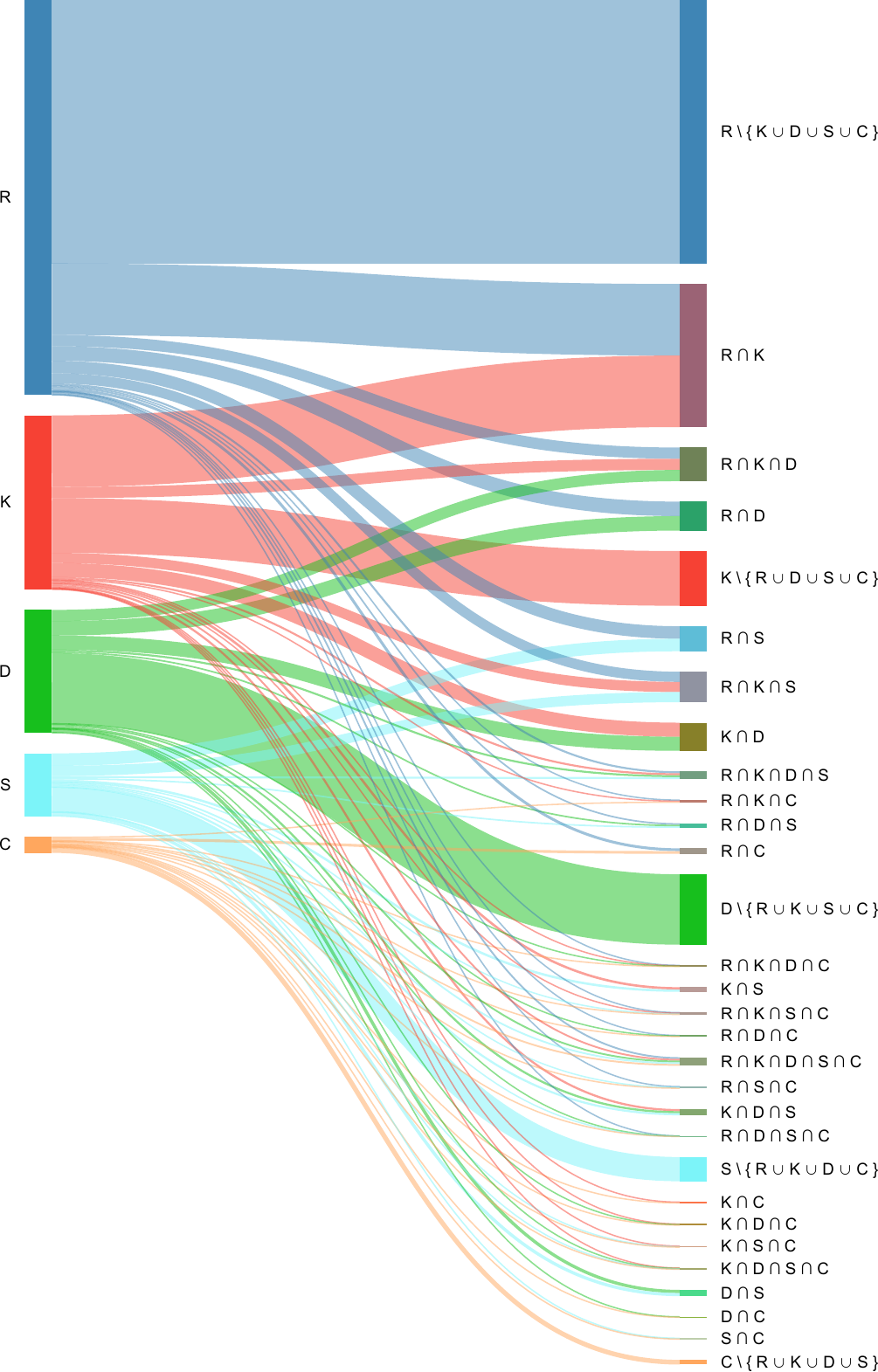}
  \caption{\label{figure:database_category_associations_top_5}Sankey diagram of the 5 DB categories associations.}
\end{figure}

\tabref{table:database_category_associations} presents them formally with the mathematical notation of sets (\eg $R$, $K$), difference ($\backslash$), union ($\cup$), and intersection ($\cap$). Then, \figref{figure:database_category_associations_top_5} shows the corresponding Sankey diagram~\citewhite{schmidt2008sankey} highlighting the frequency of each pattern.

\begin{table}[!ht]
  \centering
  {\small
  \caption{DB categories associations in microservices.}\label{table:database_category_associations}
  \begin{tabular}{lr|lr}
    \toprule
    Association & \# & Association & \# \\
    \midrule
    $ R \backslash \{ K \cup D \cup S \cup C \}$  & 337
    & $ R \cap K \cap D \cap S \cap C $  & 11 \\

    $ R \cap K $  & 183
    & $ K \cap S $  & 6 \\

    $ R \cap K \cap D $  & 43
    & $ K \cap D \cap S $  & 8 \\

    $ R \cap D $  & 37
    & $ R \cap S \cap C $  & 3 \\

    $ K \backslash \{ R \cup D \cup S \cup C \} $  & 70
    & $ R \cap D \cap S \cap C $  & 1 \\

    $ R \cap K \cap S $  & 39
    &  $ K \cap C $  & 2  \\  

    $ R \cap S $  & 32
    & $ K \cap D \cap C $  & 2 \\

    $ R \cap K \cap D \cap S $  & 10
    & $ D \cap S $  & 8 \\

    $ K \cap D $  & 36
    & $ K \cap D \cap S \cap C $  & 2 \\

    $ R \cap K \cap C $  & 4
    & $ K \cap S \cap C $  & 2 \\

    $ R \cap C $  & 7 
    & $ S \backslash \{ R \cup K \cup D \cup C \} $  & 31 \\

    $ R \cap K \cap D \cap C $  & 2
    & $ D \cap C $  & 1 \\

    $ R \cap D \cap S $  & 6
    & $ S \cap C $  & 2 \\

    $ D \backslash \{ R \cup K \cup S \cup C \} $  & 90
    & $ C \backslash \{ R \cup K \cup D \cup S \} $  & 6 \\

    $ R \cap D \cap C $  & 2
    & $ D \cap S \cap C $  & 0 \\

    $ R \cap K \cap S \cap C $  & 3
    & & \\

    \bottomrule
\end{tabular}}
\\
\medskip
{\raggedbottom \scriptsize Sets: (R)elational = 720, (K)ey-Value = 423, (D)ocument = 259, (S)earch = 164, (C)olumn = 50}.
\end{table}

The most popular association patterns for two categories are: $\langle$Relational, Key-Value$\rangle$ (18,21\%), $\langle$Relational, Document$\rangle$, and $\langle$Rela\-tional, Search$\rangle$. For triplets: $\langle$Relational, Key-Value, Document$\rangle$ (4,28\%) and $\langle$Relational, Key-Value, Search$\rangle$. On the side of quartets, $\langle$Relational, Key-Value, Document, Search$\rangle$ is the most popular one with 10 occurrences, even though it represents only 1\% of the dataset. For quintets, we reveal 11 microservices repositories opting for $\langle$Relational, Key-Value, Document, Search, Column$\rangle$. Once again, this happens only for a tiny fraction of the repositories in the collected dataset, but the fact that there are a few repositories mixing all the categories highlights the high variety \emph{in the wild}.

The relationships between the associations are not transitive. For example, no repository follows the $\langle$Document, Search, Column$\rangle$ pattern, although $\langle$Document, Search$\rangle$, $\langle$Document, Column$\rangle$, and $\langle$Search, Column$\rangle$ associations exist. This last duo also shows that in the top 5, some repositories exist without any DB categories from the top 3. Furthermore, excluding only the top category (Relational DBs), we can observe that several associations exist that do not include the most popular category. For instance, we can notice 36 microservices repositories with only Key-Value and Document DBs. Another interesting observation is the number of microservices repositories containing one and only one category of DBs. For instance, 337 out of 1,005 repositories (33.53\%) contain only Relational DBs, 90 (8.96\%) contain only Document DBs, and 70 (6.97\%) contain only Key-Value DBs. Those are in the complement of the 0.47 DHR.

\begin{figure}[!ht]
  \centering
  \includegraphics[width=\linewidth]{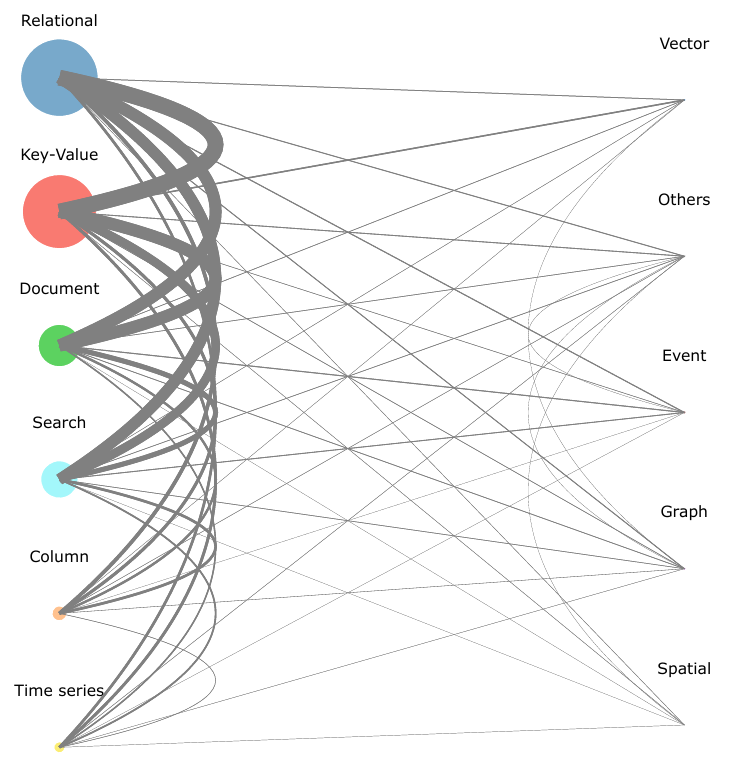}
  \caption{\label{figure:database_categories_association_dual_mainstream_vs_specific}Pairwise associations between mainstream (left) and niche (right) DB categories.}
\end{figure}

To analyze how niche DBs are connected with mainstream ones, we propose a graph-based representation depicting links between DB categories in \figref{figure:database_categories_association_dual_mainstream_vs_specific}.

Nodes represent DB categories, whose size depends on the popularity, and links represent DB associations, where the size of the link is proportional to the number of associations in that category. Nodes on the left represent mainstream DB categories, while nodes on the right represent niche ones.

In line with previous observations, this graph confirms that niche DBs are rarely associated with each other (0.37\%). It also shows that, in most cases, niche DBs are commonly associated with a mainstream one (12.34\%). Main\-stream-mainstream DB associations are the most popular (87.29\%).

Finally, considering specific DB technologies, the most popular associations are $\langle$PostgreSQL, Redis$\rangle$, $\langle$Redis, MongoDB$\rangle$, and $\langle$PostgreSQL, MongoDB$\rangle$ for duos, $\langle$PostgreSQL, Redis, Elastic\-search$\rangle$ and $\langle$PostgreSQL, Redis, MongoDB$\rangle$ for trios, and $\langle$Postgre\-SQL, Redis, MongoDB, Elasticsearch$\rangle$ for quartets.

\medskip
\InsightBox{
  \smallskip
	\textbf{\emph{RQ2: Findings}}
	\smallskip
  \begin{itemize}
    \item[F2.1] \emph{52\% of the repositories declare two DB technologies and 47\% contain DBs from two distinct categories. Half of the repositories are mono-technology/-category.}
    \item[F2.2] \emph{Some microservices repositories also declare several DBs belonging to the same DB category.}
    \item[F2.3] \emph{The DB categories practitioners combine the most (in pairs, trios, and all together) are Relational, Key-Value, Document, and Search, associated according to several different and non-transitive patterns.}
    \item[F2.4] \emph{There are empty association patterns (\eg Document, Search, and Column) that should be investigated further.}
    \item[F2.5] \emph{In most cases, niche DBs are associated with a mainstream one (12.34\%) and rarely with other niche DBs (0.37\%).}
    \item[F2.6] \emph{The most popular DB technology associations are Postgre\-SQL with Redis for duos, Postgre\-SQL, Redis, Elastic\-search for trios, and Postgre\-SQL, Redis, MongoDB, and Elastic\-search for quartets.}
  \end{itemize}
}

\subsection{RQ3: Microservices Complexity \& Databases} \label{section:result_rq3}

\begin{figure}[ht]
\centering
\includegraphics[width=0.6\linewidth]{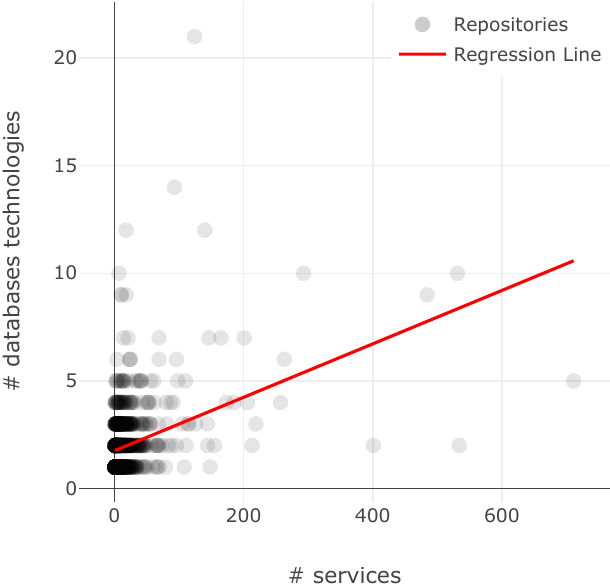}
    \caption{\label{figure:microservices_services_vs_databases}Comparison of the number of services and the number of DB technologies in microservices.}
\end{figure}

\begin{figure}[ht]
\centering
\includegraphics[width=0.6\linewidth]{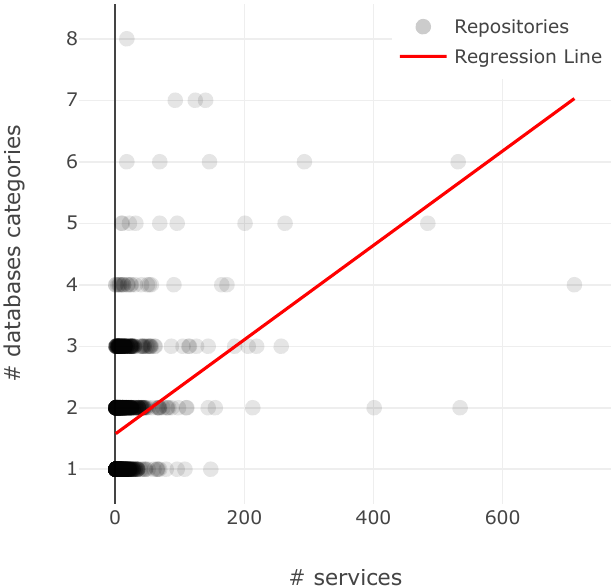}
    \caption{\label{figure:microservices_services_vs_databases_categories}Comparison of the number of services and the number of DB categories in microservices.}
\end{figure}

\begin{figure}[ht]
\centering
\includegraphics[width=0.6\linewidth]{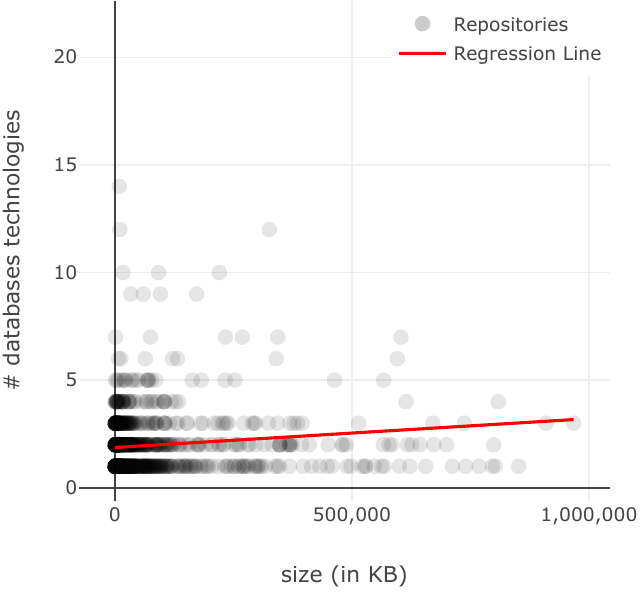}
    \caption{\label{figure:microservices_size_vs_databases}Comparison of the repository size on disk and the number of DB technologies in microservices.}
\end{figure}

\begin{figure}[ht]
\centering
\includegraphics[width=0.6\linewidth]{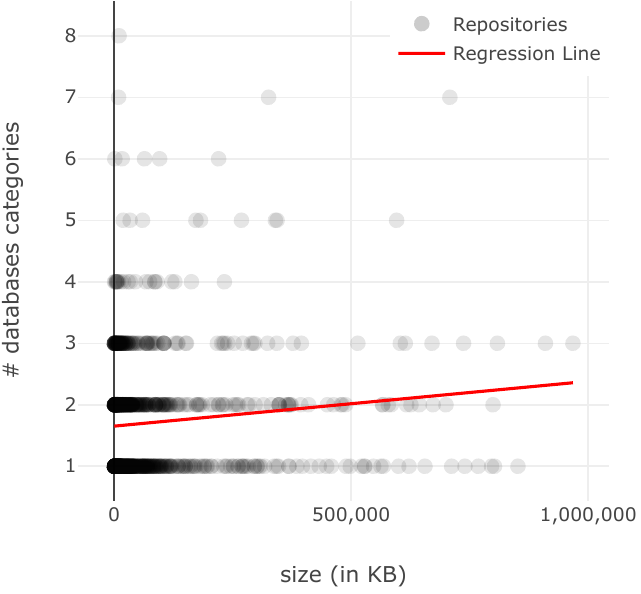}
    \caption{\label{figure:microservices_size_vs_databases_categories}Comparison of the repository size on disk and the number of DBs categories in microservices}
\end{figure}

To analyze the complexity of microservices and their DBs, we compute the number of services (excluding DBs) declared in the Docker Compose file(s). We create scatter plots comparing the number of services (x-axis) with the number of DB technologies, in \figref{figure:microservices_services_vs_databases}, and categories, in \figref{figure:microservices_services_vs_databases_categories}, on the y-axis. We use the number of services within a microservices architecture as a proxy measure of complexity. We estimate the slope of the linear regression and plot the corresponding regression lines to identify trends and explore the potential relationship between microservice complexity (x-axis) and DB (y-axis). 

We also perform a Student's t-test to assess whether the slope differs significantly from zero. The null hypothesis assumes no linear relationship (\ie a slope of zero). We consider the null hypothesis rejected if the resulting $p$-value is less than or equal to 0.05, indicating statistical significance. In our case, the linear regressions suggest that the more services there are, the more DBs there are. Results are confirmed as statistically significant ($p$-value $\leq 0.05$).

Following the same approach, we analyze another perspective with a different complexity proxy. We compare the disk size of each repository (x-axis) with the number of DB technologies in \figref{figure:microservices_size_vs_databases} and categories in \figref{figure:microservices_size_vs_databases_categories} (y-axis). Larger microservices architectures tend to have more DBs in a statistically significant way ($p$-value $\leq 0.05$).

The two complexity proxies we used agree on indicating the complexity of a microservices architecture as linked to the number of DBs it contains.

To complement this observation with concrete examples, we analyze the popular DB technologies and categories in projects that the considered proxies identify as complex. We select 61 projects among the 1,005 that have at least 80 MB in size, 20 services, and 2 DBs.

The popular technologies differ slightly from those found across the entire dataset (see RQ1). The top 5 are Redis, PostgreSQL, MySQL, Elasticsearch, and MongoDB.

Redis overtakes PostgreSQL as the most used DB, indicating a shift from Relational to Key-Value in complex architectures, likely due to increased caching needs. Elasticsearch surpasses MongoDB, highlighting the growing role of Search DBs in complex scenarios.

At the category level, the distribution remains similar to the overall dataset, with Relational DBs still offering the most diverse technology options.

\medskip
\InsightBox{
  \smallskip
	\textbf{\emph{RQ3: Findings}}
	\smallskip
  \begin{itemize}
    \item[F3.1] \emph{The complexity of a microservices architecture in terms of number of services is correlated both with the number of DB technologies and categories.}
    \item[F3.2] \emph{The number of DBs and DB categories is dependent from the size of microservices architectures.}
    \item[F3.3] \emph{Relational DBs are still prevalent, regardless of the complexity of microservices architectures.}
    \item[F3.4] \emph{In the most complex systems, DB technologies are in a different order of popularity than in the complete dataset. Redis is more popular than Postgres in complex systems and Elasticsearch is above MongoDB.}
   \end{itemize}
}

\subsection{RQ4: Microservices Age and Databases} \label{section:result_rq4}

\begin{figure}[ht]
\centering
\includegraphics[width=0.6\linewidth]{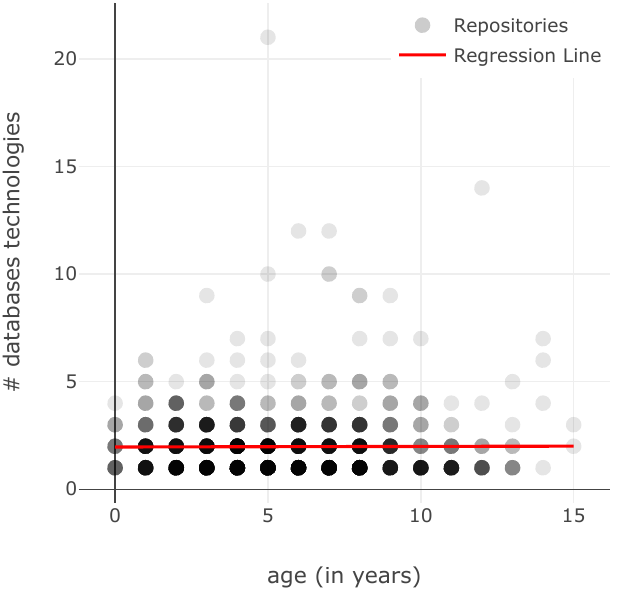}
  \caption{\label{figure:microservices_age_vs_databases}Age vs. number of DB technologies.}
\end{figure}

\begin{figure}[ht]
\centering
\includegraphics[width=0.6\linewidth]{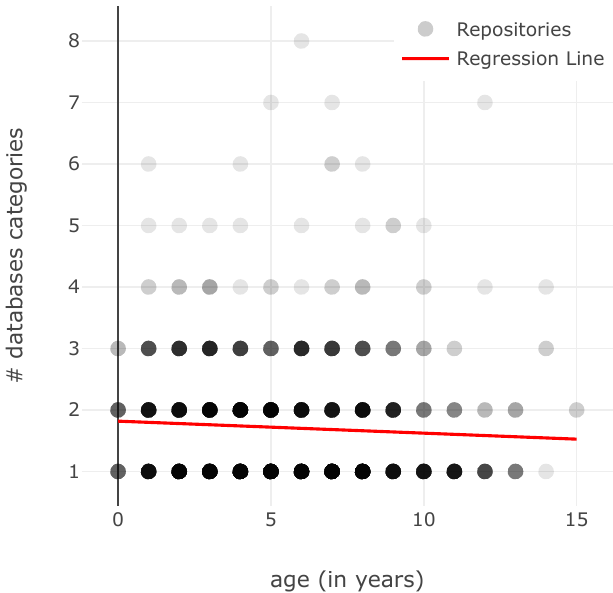}
  \caption{\label{figure:microservices_age_vs_database_categories}Age vs. number of DB categories.}
\end{figure}

The results presented in the following scatter plots show the relation between the age of a repository (x-axis) and its number of DB technologies in \figref{figure:microservices_age_vs_databases} and categories in \figref{figure:microservices_age_vs_database_categories} (y-axis).

We estimate the slope, draw the regression lines, and perform the Student's t-test to derive the $p$-value. Results indicate that we cannot reject the null hypothesis and thus we cannot conclude statistically significant correlations ($p$-value~$> 0.05$) between the age of the repositories and the number of DB technologies and categories used.

We conduct complementary investigations to identify the 5 most popular DB technologies and categories according to two age groups.

We analyze the oldest (13 years or more) and the most recent (2 years or less) microservices we collected. PostgreSQL, MySQL, Redis, and MongoDB are the most used in both age groups (although in a different order). Among the oldest, the top 3 are Relational. In contrast, we observe a more diverse distribution in newer projects, with the introduction of Key-Value and Document DBs. While MariaDB is more popular among older projects, Elasticsearch has taken its place in the top 5 for newer projects, confirming a clear shift.

\medskip
\InsightBox{
  \smallskip
	\textbf{\emph{RQ4: Findings}}
	\smallskip
  \begin{itemize}
    \item[F4.1] \emph{No conclusion can be drawn regarding a correlation between the age of a microservices architecture and the heterogeneity of its DBs.}
    \item[F4.2] \emph{Microservices shifted from Relational DBs of older architectures (PostgreSQL, MySQL, and MariaDB as the top 3 technologies), to Key-value, Document and Search categories, with Redis, MongoDB, and Elasticsearch leading in popularity in younger projects (less than 2 years old) after PostgreSQL.}
    \item[F4.3] \emph{The relational model remains predominant although it is now 5 times less present in the most recent microservices.}
    \end{itemize}
}

\section{Implications and Recommendations} \label{section:discussion}

In this section we analyze the implications of the presented results for the state of practice. We compare our findings with previous literature to highlight up-to-date recommendations for practitioners.

\subsection{Database Usage and Prevalence in Microservices}

Relational and Document DB categories are expected to be prevalent in microservices due to their popularity in monolithic architectures and their ability to manage structured and semi-structured data~\citewhite{laigner2021,benats2021}. Document and Key-Value DBs are common due to their flexibility and adaptability to the dynamic nature of microservices~\citewhite{laigner2021}. Search DBs are also expected to be used, fulfilling a specific role in microservices that require search capabilities~\citewhite{laigner2021,benats2021}. We confirm the popularity of these four categories, noting that Key-Value DBs are way more popular (42.09\%) than Document DBs (25.77\%). This is a key difference with respect to what has been found for general software~\citewhite{benats2021}. Our work also highlights how Column and Time Series DBs are important for microservices architectures, given the fact that these categories are present in tens of repositories (respectively, ranking 5th and 6th in the top 10).

In terms of technologies, PostgreSQL, MySQL, SQL Server, Redis, MongoDB, and Elasticsearch are expected to be among the most commonly used DB technologies~\citewhite{laigner2021,benats2021}. Our results confirm their popularity, now reporting them in a ranking dedicated to microservices.

Adding to this picture, MariaDB, etcd, Cassandra, and InfluxDB are also part of the top 10 most popular technologies, while previously they were flying under the radar. There are ``safe'' choices that, by general consensus, are preferable. We can refer to the top technology in each category including PostgreSQL for Relational DBs, Redis for Key-Value, MongoDB for Document, Elasticsearch for Search, Cassandra for Column, InfluxDB for Time Series, Milvus for Vector, EventStoreDB for Event, Neo4j for Graph, and PostGIS for Spatial DBs.

Regarding less common and specialized DB technologies, they are expected to address specific requirements within microservices, as encouraged by the architectural style~\citewhite{laigner2021}. Our results reveal the specific DB technologies selected for such niche goals. Examples include EventStoreDB, Neo4j, and PostGIS.

\medskip
\InsightBox{
	\smallskip
  \begin{itemize}
    \item[R1.1] \textbf{Variety training}. \emph{Microservices practitioners should be aware of the variety of DB categories and technologies commonly used in such architectures. Training on Relational, Key-Value, Document, and Search DBs is still essential.}
    \item[R1.2] \textbf{Niche awareness}. \emph{Microservices developers should be aware of specific DB categories addressing niche goals, such as Column, Time Series, Vector, Event, Graph, or Spatial DBs.}
    \item[R1.3] \textbf{Popularity consideration}. \emph{Microservices developers' training should include popular technologies like PostgreSQL, MongoDB, Elasticsearch, and eventually Cassandra, InfluxDB, Milvus, EventStoreDB, Neo4j, and PostGIS.}
    \item[R1.4] \textbf{Technology vs. principles}. \emph{In some categories (\eg Event, Graph, Spatial), knowledge of the most popular technology is enough to guarantee direct applicability in most cases, while for other (\eg Relational), knowledge of the underlying principles can improve ``knowledge portability'' with respect to the many alternative technologies.}
  \end{itemize}
}

\subsection{Database Associations in Microservices}

Regarding DB associations in microservices, practitioners often combine multiple technologies and typically at least two distinct categories~\citewhite{laigner2021}. Our results confirm this degree of heterogeneity on technologies. 52\% of the selected repositories declare two DB technologies, with 47\% from two distinct categories. Nevertheless, the remaining half does not embrace polyglot persistence, declaring only a single technology and category. Surprisingly, some microservices repositories declare several DBs belonging to the same category. These are interesting candidates for future studies on the role and evolution of co-existing technologies that our dataset provides.

The association of Relational and Document DBs together has been previously found to be prevalent~\citewhite{laigner2021,benats2021}. Some repositories also integrate with them Key-Value and Search-based DBs, all simultaneously~\citewhite{laigner2021,benats2021}.

We pointed out the various patterns in the associations among these categories. Other associations with niche DBs concern only a few microservices. While DB categories are more commonly combined in pairs or trios, there exist repositories without any Relational, Key-Value, and Document DBs (\ie the top 3 categories). There is an interesting and unexpectedly empty association pattern concerning the combination of Document, Search, and Column categories that deserves further attention.

In the literature, associations among PostgreSQL, MongoDB, Redis, and Elasticsearch, have been found to occur more frequently~\citewhite{laigner2021,benats2021}. Our analyses confirm this result. Instead, our analyses highlight how niche categories (\eg Vector, Event, Graph, Spatial) are less commonly combined with each other. Practitioners prefer to pair them with mainstream ones.

\medskip
\InsightBox{
	\smallskip
  \begin{itemize}
    \item[R2.1] \textbf{Heterogeneity handling}. \emph{Microservices practitioners should handle the heterogeneity and association of DB technologies, especially among Relational, Key-Value, Document, and Search DBs, by being able to master two or three categories at the same time. Competence with a single specific DB is often insufficient.}
    \item[R2.2] \textbf{Homogeneity acceptance}. \emph{However, in about half of the cases, homogeneity is observed. Microservices architectures developers should also be prepared for single-category and single-technology scenarios, regardless of the polyglot persistence.}
  \end{itemize}
}

\subsection{Microservices Complexity and Databases}

Microservices tend to increase the technical complexity~\citewhite{assunccao2023}. It is expected to positively correlate with the use of a greater number of DB technologies~\citewhite{gan2019b,soldani2018}. For instance, microservices network centrality derived from inter-service calls is associated with the number of public methods and call frequency~\citewhite{bakhtin2025}, reflecting the potential increase in data access and, consequently, in the number of DBs. Our results confirm that the number of services and the size of the project are correlated with the number of DB categories and technologies.

In terms of technologies, complex architectures should favor Document DBs for their flexibility in handling constraints and Key-Value DBs for caching and performance optimization~\citewhite{laigner2021}. We found that in the most complex systems in our dataset, Redis is over PostgreSQL and Elasticsearch is over MongoDB. Nevertheless, Relational DBs remain very popular with multiple alternatives. As a side effect, this gives to Redis the top spot among the technologies.

\medskip
\InsightBox{
  \begin{itemize}
    \item[R3.1] \textbf{Complexity management}. \emph{Practitioners contributing to complex systems should be able to manage multiple DB technologies across different categories.}
    \item[R3.2] \textbf{Variety-induced gap}. \emph{Although Relational DBs are prevalent in general, the single most relevant technology for practitioners is Redis.}
  \end{itemize}
}

\subsection{Microservices Age and Databases}

We analyzed the relationships between the age of the architecture and number of DB categories and technologies. Older microservices are expected to accumulate more DBs~\citewhite{gan2019b,soldani2018}. Our analyses show there is no statistically significant correlation between the two metrics.

Regarding the popularity of specific categories over time, trends suggest a slightly decreased reliance on Relational DBs~\citewhite{benats2021}, \citegrey{DBEnginesTrendsHistoryRelational}, especially in older microservices architectures refactoring their codebase to leverage the newer technologies.

Our results confirm this scenario, with Relational DBs popular in older microservices (13 years or more) while Document, Column, and Search DBs, are preferred in recent systems~\citewhite{laigner2021,benats2021}, \citegrey{DBEnginesTrendsHistoryDocument,DBEnginesTrendsHistoryKeyValue,DBEnginesTrendsColumnSearch,DBEnginesTrendsHistorySearch}.

\medskip
\InsightBox{
\begin{itemize}
	\item[R4.1] \textbf{Cautious replacement}. \emph{Microservices contributors should not discard, over time, Relational DBs in favor of other ``modern replacement'' DBs, given their deep-rooted popularity, even in microservices.}
\end{itemize}
}
\section{Threats to Validity}\label{section:threats_validity}

In this section, we discuss limitations and threats that may affect the validity of our study and how we mitigated them.

\subsection{Construct validity}

Threats to construct validity concern the relation between theory and observation. Our identification of microservice repositories relied on heuristics such as keywords, programming languages, and structural indicators. While these heuristics were carefully designed, some false positives and negatives may persist, potentially including non-microservices and excluding relevant repositories that do not explicitly match some of the criteria.

To partially mitigate this threat, multiple conditions must be satisfied for the repository to be included in the dataset, with more emphasis on precision rather than recall, ensuring the quality of our dataset as future benchmark.

Another threat is the implicit exclusion of DB technologies that lack Docker images, as the presence of Docker files was used as a distinguishing criterion. Although these exclusions are limited and do not involve the most popular DBs, they may still slightly affect the comprehensiveness of our results. The historical perspective presented in RQ4 is based on a fixed snapshot and considers only the age of the repository computed from the creation date. While this offers initial insights, a more fine-grained analysis of the actual evolution history of repositories would provide a better understanding of trends.

Finally, our analyses considered all DBs declared in Docker Compose files, regardless of whether they were actively used in the codebase. This could lead to the inclusion of unused or ``ghost'' dependencies. A more in-depth approach, for example leveraging static program analysis, would be necessary to confirm the actual usage of declared DBs in the application code.

\subsection{Internal validity}

Internal validity concerns how one can be confident on claimed cause and effect relation. We do not claim any causation in our study. We analyze the (joint) usage of DB technologies in microservices applications, and assess possible correlations with system complexity and age. Hence, this study is not subjected to threats to internal validity.

\subsection{External validity}

External validity concerns the generalizability of findings beyond the study context. Our study considers different types of projects in terms of application domain, size, complexity, programming languages, and DB technologies. Although we ensured to collect an heterogeneous sample with respect to these criteria, we only considered \emph{open-source} projects and DB technologies in public GitHub repositories. These choices affect the generalizability of our findings, yet our approach remains valid and the insights pertinent in the context of industrial software systems leveraging the same DBs.

\subsection{Conclusion validity}

Threats to conclusion validity concern the degree to which the statistical conclusions about the claimed relationships are reasonable. To reduce bias in results interpretation, we used standard statistical methods such as linear regression, Student's t-test, and $p$-values.

\subsection{Reliability validity}

Reliability validity concerns factors that could cause an error in data collection and analysis. To minimize potential threats to reliability, we analyzed open-source projects publicly available on GitHub and provided a replication package that contains our dataset and all the scripts we used for our analyses.

\section{Related Work} \label{section:related_works}

Most related studies focus on DBs \emph{or} on microservices, but \emph{separately}. In this section, we review the relevant literature to contextualize and position our work, emphasizing its contribution to bridging these two domains.

\subsection{Databases}


Curino \etal~\citewhite{curino2008} propose an empirical study on DB schema changes in real-world data-intensive open-source projects, focusing on Wikipedia as a case study. They highlight the challenges of maintaining and evolving such relational DBs and provide observations based on 4.5 years or history and 171 revisions, culminating in 34 tables, 242 columns, and 700GB of data. Their goal is to illustrate typical evolution scenarios through types of changes, warn about common design errors, and offer recommendations to researchers. This work represents a step towards a unified benchmark for researchers built on a real, complex, and large case study. 

\newpage

In the same vein, Qiu \etal~\citewhite{qiu2013} empirically analyze the co-evolution of relational DB schemas and related source code in 10 popular and large projects composed of over 160K revisions collected from the Subversion version control system. They demonstrate the high frequency of DB changes in the software life cycle, their impact on the source code, and the types and patterns of these changes, proposing a list of DB schema change types. 

Still focusing on relational DB schemas, Vassiliadis~\citewhite{vassiliadis2021} studies the evolution profiles, \ie the recurring activity patterns in relational DB schema evolution. The author extracts and analyzes the schemas of 195 open-source projects from \emph{Libraries.io}. This work provides practitioners and researchers with insights into evolution patterns to help them understand and predict such phenomena. 

Goeminne and Mens~\citewhite{goeminne2015} investigate the technical (co-)usage of DB access frameworks and object-relational mappers. They empirically analyze usage implications for 5 relational DB frameworks in 3.7K open-source projects in GitHub. The authors perform a survival analysis, observing combinations and complementarities across frameworks, especially those involving JDBC. They report that some technologies (\eg JPA, Spring) exhibit better survival rates. 

Decan \etal~\citewhite{decan2017} propose another empirical study on the use of relational DB access technologies like JDBC, Hibernate, and JPA in about 2.5K open-source Java GitHub projects. They perform a fine-grained analysis at file level, assessing technologies breakdown and the impact of their replacements on source code. 

Small to medium scale case studies fail to capture general trends in a large, {\em in-vivo} population of real world applications. Large studies, on the contrary, focus on a limited amount of technologies, often in a single category, often relational, lacking qualities of a broad spectrum overview like in our study.


On the NoSQL side, Gessert \etal~\citewhite{gessert2017} compare several technologies, particularly for Key-Value, Document, and Column store categories. They provide practitioners with a decision tree to support choices based on functional and non-functional requirements. 

Davoudian \etal~\citewhite{davoudian2018} present a comprehensive survey on various NoSQL technologies, including Key-Value, Column, Document, and Graph DBs. They analyze these technologies from the perspectives of data models, consistency models, data partitioning strategies, and the CAP theorem (Consistency, Availability, and Partition tolerance). Their work includes both academic and industrial examples, providing valuable insights to assist practitioners in making informed decisions on which technology to use. 

Scherzinger and Sidortschuk~\citewhite{scherzinger2020} present an empirical study on NoSQL DBs, analyzing 1.2K open-source Java projects and their GitHub history. They confirm common practices in schema-free data modeling and evolution scenarios. 

These work focus on the new type of NoSQL DB categories but lack the context of the preceding and still surviving technologies. As confirmed by our findings on the collected dataset, including both SQL and NoSQL categories and technologies, the former are still far from being overthrown or obsolete.

\newpage


A first step in the direction of a comparison between old and new is the work by Benats \etal~\citewhite{benats2021}, unifying relational and NoSQL data models in an empirical study investigating the use of hybrid multi-DBs over time in different languages. They consider 4 years of history across over 40K open-source projects from {\em Libraries.io}. We compare their empirical study with our findings (\secref{section:discussion}). 

More recently, Paiva \etal~\citewhite{paiva2025} proposed an empirical study encompassing all data models in 362 Java open-source projects from GitHub. Studying the popularity and combinations of DB technologies, their stability, migration patterns, and the role of object-relational mappers, they provide insights to researchers and practitioners for selecting appropriate DB technologies. 


In the mobile domain, Lyu \etal~\citewhite{lyu2017} investigate local DBs for Android by conducting an empirical study on 1,000 popular apps from the Google Play Store. They provide an overview of available technologies and their usage, identifying major problems and deriving recommendations for developers. 

Our work compares the findings in microservices architectures with the ones in generic software systems at large, highlighting similarities and differences specific to the microservices domain.

\subsection{Microservices}

Brogi \etal~\citewhite{brogi2017} argue that previous works lack a reference dataset of open-source microservices projects, like a standardized benchmark. Thus, they propose \textit{µ}\textsc{set}, 5 microservices projects easy to set up for repeatable experiments. These projects are developed {\em ad hoc}, according to 5 common requirements in microservices, as identified by the authors. 

Rahman \etal~\citewhite{rahman2019} propose the first dataset including real-world microservices open-source projects from GitHub. This list of 62 projects contains monoliths migrated to microservices or projects developed from scratch following the microservices architecture. 

In a follow-up work, d'Aragona \etal~\citewhite{dAragona2024} enlarge this dataset with 378 open-source projects from World of Code, developed in several languages. They document each project with additional data and insights helping researchers to select the most appropriate items for their work. In comparison to our work, the authors deliberately exclude the DBs from their study, focusing only on the modular nature of the microservices architectures of those systems. 

Finally, Wang \etal~\citewhite{yang2024} present a dataset of microservices applications utilizing Spring Cloud. Their contribution aims to complete previous works by suggesting complex fine-grained metrics in order to understand bad code smells in microservices. 

\subsection{Microservices and Databases}

Only works by Gan and Delimitrou~\citewhite{gan2018}, Gan \etal~\citewhite{gan2019}, and Laigner \etal~\citewhite{laigner2024} propose, as benchmarks for researchers, microservices architectures repositories written in various popular languages and with different DBs. They aim to provide standard baselines for their studies and subsequent research. In all these cases, the intended benchmarks are comprised of a single to a maximum of six {\em ``synthetic and prototypical''} end-to-end applications.

To the best of our knowledge, our empirical study is the first to bring together over 1,000 open-source microservices projects, in 11 different languages, to investigate their use of DBs across a wide variety of technologies and categories, surpassing the previous works discussed above in either size of the dataset or number and variety of considered DB technologies and categories.


\section{Conclusion} \label{section:conclusion}

We presented an empirical study on DB usage in microservices, analyzing one thousand open-source projects from GitHub developed in the last 15 years.

Our work addresses questions regarding the prevalence of DB categories and technologies used in microservices. We investigated, from several perspectives, the way DBs are combined in practice, observing recurrent patterns. We deepened our observations with objective fine-grained metrics and highlight relationships between different characteristics (\eg complexity vs. age).

Besides the ``usual suspects'', we shed light on less common DBs like Time Series, Vector, Event, Graph, and Spatial DBs addressing niche goals. We highlighted a variety of Relational technologies and, overall, a variety of DBs with up to 60 unique technologies identified.

We show how microservices rely on heterogeneous DBs. Half of them use multiple DB technologies across different DB categories. Consequently, the other half uses a single technology and a single category, with unclear implications on the best strategy for database practitioners to prepare for a career path involving microservices architectures. From our analyses it emerges the large spectrum of combination patterns. Nevertheless, we try to pinpoint findings leading to practical observations and recommendations for practitioners. The 18 findings and 9 recommendations we derive are the simplest yet faceted representation of such a complex and heterogeneous reality.

Finally, we emphasize that larger microservices architectures tend to leverage more and diverse DBs. We analyze how still leading Relational DBs, shifting across the years, are now five times less prevalent in microservices, to the profit of emerging technologies such as Document, Key-Value, and Search DBs.

As a concluding remark, fundamental contributions of this work are also the dataset and the systematic approach with which we automatically built it. The published dataset is the factual basis for the reflections presented in this work and constitutes a sound starting point for future large scale research endeavors. Our work supports researchers and practitioners in understanding, evolving, and optimizing DB usage in microservices architectures.


\section*{Replication package and dataset} \label{section:replication_package_dataset}

To ensure transparency, verifiability, and reproducibility of our work, all the artifacts resulting from our study are available at: \faFileArchive[regular]~{\urltt{https://github.com/DatabaseEvolutionNudgeInMicroservices/daim}}

This repository includes the MongoDB database with our complete dataset (\ie the list of GitHub repositories considered) and the scripts used to perform the analyses and to generate charts, tables, matrices, and metrics in the present work.
\section*{Acknowledgments}

This work was supported by the SofinaBoël Fund for Education and Talent; the Federation Wallonie-Bruxelles (FWB), as part of the ARC project RAINDROP; and the Swiss National Science Foundation (SNSF) through the project ``FORCE'' (SNF Project No. 232141).

\bibliographystylewhite{elsarticle-num}
\bibliographywhite{bibliography}

\bibliographystylegrey{elsarticle-num}
\bibliographygrey{bibliography}

\end{document}